\documentclass[%
 reprint,
groupedaddress,
frontmatterverbose, 
showpacs,preprintnumbers,
 amsmath,amssymb,
 aps,
 pra,
floatfix
]{revtex4-1}

\usepackage{graphicx,url,graphics}
\usepackage{dcolumn}
\usepackage{bm}
\usepackage{longtable}                                                                            
\begin{document}
\title{Valence-shell single photoionization of Kr$^{+}$ ions: Experiment and Theory}

\author{G. Hinojosa}\email{hinojosa@fis.unam.mx}
\altaffiliation{Present address: Instituto de Ciencias F\'{i}sicas, Universidad NacionalAut\'onoma 
			de M\'exico, Apartado Postal 48-3, Cuernavaca 62210, Morelos, M\'exico},

\author{A. M. Covington}, \author{G.  A. Alna'Washi}\email{alnawashi@hu.edu.jo}
\altaffiliation{Present address: Department of Physics, The Hashemite University, Zarqa 13115, Jordan}
\author{M. Lu},
\author{R. A. Phaneuf}
\affiliation{Department of Physics, University of Nevada, Reno, NV 89557-0220}

\author{M. M. Sant'Anna}
\affiliation{Instituto de F\'\i sica, Universidade Federal do Rio de Janeiro, Caixa Postal 68528 - CEP, 21941-972 Rio de Janeiro RJ, Brazil}

\author{C. Cisneros}, \author{I. \'{A}lvarez}
\affiliation{Instituto de Ciencias F\'{i}sicas, Universidad Nacional
		Aut\'onoma de M\'exico, Apartado Postal 48-3, Cuernavaca 62210, Morelos, M\'exico.}

\author{A. Aguilar}, \author{A. L. D. Kilcoyne}, \author{A. S. Schlachter}
\affiliation{Advanced Light Source, Lawrence Berkeley National Laboratory, 1 Cyclotron Road, Berkeley, CA 94720}

\author{C. P. Ballance}
\affiliation{Department of Physics, 206 Allison Laboratory, Auburn University, Auburn, AL 36849-5311}

\author{B. M. McLaughlin}\email{b.mclaughlin@qub.ac.uk}
\altaffiliation[Present address: ]{Centre for Theoretical Atomic, Molecular and Optical Physics (CTAMOP), School of Mathematics and Physics,
                      The David Bates Building, 7 College Park, Queen's University of Belfast, Belfast BT7 1NN, United Kingdom}
\affiliation{Institute for Theoretical Atomic and Molecular Physics,
       		Harvard Smithsonian Center for Astrophysics, 
		60 Garden Street, MS-14, Cambridge, MA 02138}

%
%

\date{\today}

\begin{abstract}
  Photoionization of Kr$^+$ ions was studied in the energy range from 23.3 eV to 39.0 eV at
  a photon energy resolution of 7.5 meV.  Absolute measurements were performed by merging beams of Kr$^+$ ions and of
  monochromatized synchrotron undulator radiation. Photoionization (PI) of this Br-like ion is
  characterized by multiple Rydberg series of autoionizing resonances
  superimposed on a direct photoionization continuum. 
    Resonance features observed in the experimental spectra are spectroscopically assigned and their energies and 
  quantum defects tabulated.    The high-resolution cross-section measurements are benchmarked against state-of-the-art theoretical 
  cross-section calculations from the Dirac-Coulomb R-matrix method \cite{McLaughlin2012}. 
    
\end{abstract}



\maketitle
%
%
%
%

\section{Introduction}

 The identification of spectroscopic lines from ions  is crucial for studying
 practically all regions of the Universe, as most of the matter in the interstellar 
 medium is in an ionized state \cite{cartlidge2012}. 
The photoionization of krypton ions is of particular interest
 in the determination of elemental abundances in
 stars and planetary nebulae~\cite{sterling2007}, 
 as well as in inertial-confinement fusion experiments~\cite{bizau2011}.
 Injection of Kr gas has been demonstrated to mitigate disruption in magnetically confined Tokamak
 fusion plasmas~\cite{fusion2005}.

Spectroscopic lines of Kr ions  have been used as markers in nebular spectroscopy, 
 a developing field that promises to help understand evolutionary models of 
 stars \cite{sterling2011}. In planetary nebulae, the known emission line of Kr III (Kr$^{2+}$) :
 4$s^2$4$p^4$($^3$P$_{2}$) $\rightarrow$ 4$s^2$4$p^4$($^1$D$_{2}$) has been used to identify Kr which is a
 characteristic element resulting  from the process of nucleosynthesis in 
 stars \cite{sharpee2007,sneden2008,sterling2007}, where up to 6 times ionised Kr ion stages have been detected. 
 
 For some of these ions, photoionization (PI) parameters have been measured; Kr$^{+}$ \cite{bizau2011},
 Kr$^{3+}$ \cite{Lu2006a}, Kr$^{5+}$ \cite{Lu2006b}, the Kr iso-electronic 
 sequence \cite{kilbane2007}, K-shell photoionization of Kr$^{+}$ \cite{southworth2007} 
 and extreme UV radiation photoionization \cite{richter2009,meyer2010}.
    
 A significant increase of the available photon flux from third-generation
 synchrotron light sources has allowed an important technical limitation
 due to the tenuousness of ion beam targets to be surmounted for a number of
 atomic ions~\cite{West2001,Kjeldsen2006}. This development has made accurate experimental
 data on ionic structure available to benchmark state-of-the-art quantum mechanical models. 
  
 Sophisticated methods are available for  the calculation of PI cross-sections  \cite{burke1975,Burke2011,covington2011};
 high-resolution measurements are required to verify theoretical predictions \cite{sterling2008} or reveal 
 the limitations of a particular model.  High resolution Xe$^{+}$ PI data were successfully used to  benchmark the 
 Dirac-Atomic-R-matrix-Codes (DARC) \cite{McLaughlin2012}. 
  The availability of better atomic structure  for these ion stages \cite{West2001,Kjeldsen2006} has
 prompted the refinement and development of better photoionization models \cite{McLaughlin2012,sterling2011}.
 Although the agreement between theory and experiment has improved, the accurate description 
 of the PI process is still far from trivial because  it requires a detailed analysis of 
 all open and closed interfering channels.

  Relevant to this work are the pioneering PI studies of isoelectronic atomic Bromine carried out by 
 Ru$\check{\mbox{s}}\check{\mbox{c}}$\'ic {\it et al.} \cite{berko1984} and by van der Meulen {\it et al.}
 \cite{JPB2597}. Both these measurements were conducted at lower photon energy resolution than in the present experiment but
 permitted measurement and identification of autoionizing Rydberg series converging to final
 states of Br$^{+}$, 4$s^2$4$p^4$($^3$P$_{0,1,2}$, $^1$D$_2$ $^1$S$_0$) and 4$s$4$p^5$($^3$P$^{\circ}_{0,1,2}$),
 which guided the spectroscopic assignments in the present photoionization measurements of Kr$^{+}$ ions.
  
  Similar measurements of photoionization of Kr$^{+}$ were carried out by Bizau et al.~\cite{bizau2011}
 at a photon energy resolution ranging between 50~meV near the ground-state ionization threshold of 24.36 eV,
 and 100~meV at photon energies 15~eV above threshold. They were also able to perform measurements 
 in the photon energy range 23.0 -- 27.0 eV with a spectral resolution of 30 meV.
  Although their merged-beam measurements were normalized to ground-state 
  cross sections from ion trap-measurements, the resonance structure could 
  not be fully resolved and spectroscopically assigned. With that objective, 
  the present measurements were taken with spectral resolution of 7.5 meV in 
  the energy range 23.3 --  39.0~eV.
 
 The ASTRID measurements for absolute cross sections permitted
 comparison with theoretical calculations performed using the MCDF method. 
 Bizau and co-workers identified Rydberg resonance series  4$s^2$4$p^4$($^1$D$_2$, $^1$S$_0$)n$d$ 
 that converged  to the $^1$D$_2$ and $^1$S$_0$ thresholds.
 PI cross-section calculations using the Dirac-Atomic-R-matrix-Codes (DARC)  
 on this system have  recently  been performed by McLaughlin and Ballance \cite{McLaughlin2012} in the near-threshold 
 region, which showed excellent agreement with the absolute 
 PI measurements made at the ASTRID and SOLEIL synchrotron radiation facilities.
 In the present higher resolution work, it was possible to resolve, identify and to reassign 
 a wealth  of resonance structure. Theoretical results on this complex help interpret, benchmark 
 and validate experimental measurements.
 
 %
%
%
%

\section{Experiment}

 The experiment was conducted on undulator beamline 10.0.1.2 of the Advanced Light Source
 (ALS) at Lawrence Berkeley National Laboratory, 
 using the ion-photon-beam (IPB) endstation based on the merged-beams technique \cite{phaneuf1999}. 
 The experimental method has been described in detail elsewhere in previous measurements of
 photoionization cross sections for Ne$^{+}$ \cite{covington2002}, 
 CO$^{+}$ \cite{PhysRevA.66.032718} and O$^{+}$ \cite{covington2001,aguilar2003}.

\subsection{Photon beam}
 A grazing-incidence spherical-grating monochromator dispersed the light generated
 by a 10-cm period undulator, delivering a highly collimated photon beam of spatial width less
 than 1~mm and divergence less than
 0.5$^{\circ}$. The beamline delivers a maximum flux of 5$\times$10$^{12}$ photons per second
 in a bandwidth of 0.01\% at an energy of 30 eV. In the energy
 range of 23.3 to 34.0 eV a spectral resolution of 7.5 meV was selected by
 adjusting the entrance and exit slits of the monochromator.  Fig. \ref{fullrange} shows an overview 
 of the spectra obtained at the resolution of 7.5 meV over the photon energy range 23.0 to  39.0 eV.
 
The photon energy was scanned by rotating the grating and translating the exit
 slit of the monochromator, while simultaneously adjusting the undulator gap to maximize
 the beam intensity. The photon flux was measured by an absolutely calibrated silicon
 X-ray photodiode. The analog output from a precision current meter was directed to
 a voltage-to-frequency converter, which provided a normalization signal to a
 personal-computer-based data acquisition system. The photon beam was time modulated
 (mechanically chopped) at 0.5 Hz using a stepping-motor controlled paddle to separate
 photoions from background produced by stripping of the parent Kr$^+$ ion beam on residual
 gas in the ultra-high vacuum system. The photon energy scale was calibrated using
 measurements of the known O$^{+}$ ground-state ($^{4}$S$^{\circ}$) and metastable-state ($^{2}$P$^{\circ}$
 and $^{2}$D$^{\circ}$) energy thresholds~\cite{covington2001,aguilar2003}, allowing for the Doppler
 shift due to the ion motion in the laboratory frame of reference. The absolute uncertainty
 in the photon energy scale is estimated to be $\pm$30 meV.
 
\subsection{Ion beam}
 Kr$^+$ ions were produced in a hot-filament discharge ion source. The
 filament was mounted inside a small chamber filled with Kr gas
 to a constant pressure of typically 0.1 Torr.
 A metallic cap (anode) with a 1 mm diameter hole (exit) placed as a cover, allowed ions
 to be extracted from the chamber and isolated the higher pressure region
 of the ion source from the high vacuum in the extraction region. Electrons emitted by the
 filament were accelerated by a potential difference of 100 V applied between the
 filament and the anode. Electron-atom collisions at energies $\le$100 eV are the main mechanism
 for production of Kr$^+$ ions.

Kr$^{+}$ ions extracted from the ion source were accelerated to an energy of
 6.0~ keV, forming a beam that was electrostatically focused and mass-to-charge selected by
 a 60$^{\circ}$ sector magnet.

 The surrounding volume inside the exit of the ion source chamber is a region of undetermined
 high pressure where ions suffered low energy resonant collisions with neutral Kr
 and residual gas molecules.
 It is well known that in certain collision systems, the relative populations of the
 $^{2}$P$^{\circ}_{3/2}$ and $^{2}$P$^{\circ}_{1/2}$ metastable states of noble gas
 ions may be controlled by collisional quenching at thermal energies.~\cite{Adams1980}
  
To our knowledge, collisional quenching of metastable Kr$^+$($^2$P$^{\circ}_{1/2}$) has only
 been tested for collisions with CH$_4$~\cite{Kok200047}. No clear
 experimental evidence was found that such collisions actually modify the relative
 populations of the ground and metastable states. However, the strong evidence of metastable
 collisional quenching of the Xe$^+$($^2$P$^{\circ}_{1/2}$) metastable state in collisions with N$_2$O and
 CH$_4$ supports the possibility that resonant collisions of Kr$^+$ with Kr atoms or
 vacuum remanent residual gas molecules may reduce the population of the initial
 Kr$^+$($^2$P$^{\circ}_{1/2}$) metastable state.

 \subsection{Beam-beam interaction}
 
  Two cylindrical einzel lenses and a set of four-jaw slits focused and
 collimated the Kr$^+$ ion beam. With the aid of two sets of perpendicular electrostatic steering plates,
 the ion beam was positioned and directed toward a 90$^{\circ}$ electrostatic spherical-sector deflector that
 merged the well-collimated ion beam onto the axis of the counter-propagating photon beam from the
 ALS.  Typical diameter of the photon beam was 0.1 cm, the ion beam diameter 
 varied between 0.2 cm and 0.5 cm.
 
 The interaction region consisted of an electrically isolated stainless-steel-mesh cylinder to which
 an electric potential of +2 kV was applied to energy-label Kr$^{2+}$ ions
 produced within. Entrance and exit apertures accurately
 defined the effective length of the interaction region (29.4~$\pm$~0.6 cm). Two-dimensional
 intensity distributions of both beams were measured by commercial rotating-wire beam
 profile monitors installed along the merged path just upstream and downstream of the interaction region, and by
 a translating-slit scanner located at the center of the region. The profile monitors
 permitted the positions and spatial intensity profiles of the two beams to be continuously monitored
 on an oscilloscope while tuning the beams for optimum overlap. The pressure in this region was typically
 5 $\times$ 10$^{-10}$ Torr under measurement conditions.

\subsection{Separation and detection of product ions}

 Downstream of the interaction region, a 45$^{\circ}$ dipole analyzing 
 magnet demerged the beams and separated the Kr$^{2+}$ products from the 
 parent Kr$^{+}$ beam, which was collected in an extended Faraday cup. 
 The magnetic field was set such that the Kr$^{2+}$ product ions
 passed through an aperture in the back of the Faraday cup. A spherical 
 90$^{\circ}$ electrostatic deflector directed them onto a stainless steel 
 plate biased at -550 V, from which secondary electrons were accelerated and
 detected by a microsphere-plate electron multiplier operated in pulse-counting 
 mode.

 The detection planes of the demerger magnet and this spherical detector 
 are orthogonal, permitting the Kr$^{2+}$ products to be swept across the 
 detector in mutually perpendicular directions, providing a diagnostic of 
 their complete collection. The absolute efficiency of the photoion detector 
 was calibrated {\it in situ} using an averaging sub-femtoampere meter to
 record the Kr$^{2+}$ photoion current, which was then compared to the 
 measured photoion count rate. The primary Kr$^{+}$ ion beam current was 
 measured by a precision current meter, whose analog output was directed to a 
 voltage-to-frequency converter, providing a normalization signal to the data 
 acquisition system.

%
%
%
%
%

\section{Theory}
 For comparison with high-resolution measurements  made at the
ALS, state-of-the-art theoretical methods using
highly correlated wavefunctions  are required that include relativistic effects
as fine-structure effects may be resolved.  
 An efficient parallel version \cite{ballance06}  of the DARC
  \cite{norrington87,norrington91,norrington04,darc} suite 
  of codes has been developed \cite{venesa2012,Ballance2012,McLaughlin2012} 
  to address the challenge of electron and photon interactions with atomic systems
  catering for hundreds of levels and thousands of scattering channels.
Metastable states  are populated in the Kr$^+$ ion beam  and require
additional theoretical calculations  to be carried out. 
Modifications to the Dirac-Atomic-R-matrix-Codes 
(DARC)~\cite{venesa2012,McLaughlin2012,Ballance2012} allowed high quality 
photoionization cross section calculations to be made on heavy complex systems of prime interest to 
astrophysics and plasma applications.  Cross-section calculations for trans-Fe element single photoionization of Se$^{+}$ 
and Xe$^{+}$ ions along with low ionization stages of W ions \cite{Ballance2012,McLaughlin2012,Muller2012} have been made. 

Photoionization cross sections on this halogen-like ion were 
performed for the ground and the excited metastable levels 
associated with the $4s^24p^5$ configuration for Kr$^+$ 
to benchmark the theoretical results and validate the present high resolution experimental 
measurements made at the Advanced Light Source radiation facility in Berkeley.
Details of the atomic structure calculations with the GRASP code can be found 
 in our recent work on this complex ion \cite{McLaughlin2012} where a detailed comparison 
 was made for the PI cross sections with the measurements performed at the ASTRID/SOLEIL radiation facilities.
  In the photoionization cross-section calculations for this complex trans-Fe element 
  all 326 levels arising from the seven configurations: 
 $4s^24p^4$, $4s4p^5$, $4s^24p^34d$, $4s^24p^24d^2$, $4p^6$,
 $4s4p^44d$ and $4p^44d^2$ were included  in the close-coupling expansion.   
 PI cross section calculations with this 326-level model were performed in 
 the Dirac-Coulomb approximation using the DARC codes \cite{Ballance2012,McLaughlin2012}.  

The R-matrix boundary radius of 7.44 Bohr radii  was sufficient to envelop
 the radial extent of all the $n$=4 atomic orbitals of the residual Kr$^{2+}$ ion. A basis of 16 continuum
 orbitals was sufficient to span the incident experimental photon energy
 range from threshold  up to 40 eV. Since dipole selection rules apply, 
 total ground-state photoionization require only  the 
 bound-free dipole matrices, $\rm J^{\pi}=3/2^{\circ} \rightarrow J^{{\prime}^{\pi^{\prime}}}=1/2^{e},3/2^{e},5/2^{e}$. 
 For the excited metastable states only the $\rm J^{\pi}=1/2^{\circ} \rightarrow J^{{\prime}^{\pi^{\prime}}}=1/2^{e},3/2^{e}$ are necessary.

 For both the ground and metastable initial states, the outer region electron-ion collision 
problem was solved (in the resonance region below and
 between all thresholds) using a suitably chosen fine
energy mesh of 5$\times$10$^{-8}$ Rydbergs ($\approx$ 0.68 $\mu$eV) 
to fully resolve all the extremely narrow resonance structure in the appropriate photoionization cross sections. 
The $jj$-coupled Hamiltonian diagonal matrices were adjusted so that the theoretical term
energies matched the recommended experimental values of NIST \cite{Ralchenko2010}. We note that this energy
adjustment ensures better positioning of certain resonances relative to all thresholds included in
the calculation \cite{McLaughlin2012}.  Further details of the theoretical calculations 
  can be found in our recent work on this complex ion \cite{McLaughlin2012}, 
  using the modified DARC codes, that showed suitable agreement with the 
  experimental data of Bizau and co-workers \cite{bizau2011}.
 
 In the present work the DARC PI cross-section calculations were convoluted with a Gaussian of 7.5 meV FWHM and statistically weighted
 for the ground  and metastable states to compare directly with the higher-resolution PI  measurements performed
 at the Advanced Light Source synchrotron radiation facility in Berkeley.

%
%
 
\begin{figure*}
\begin{center}
 \includegraphics[scale=1.5,width=\textwidth]{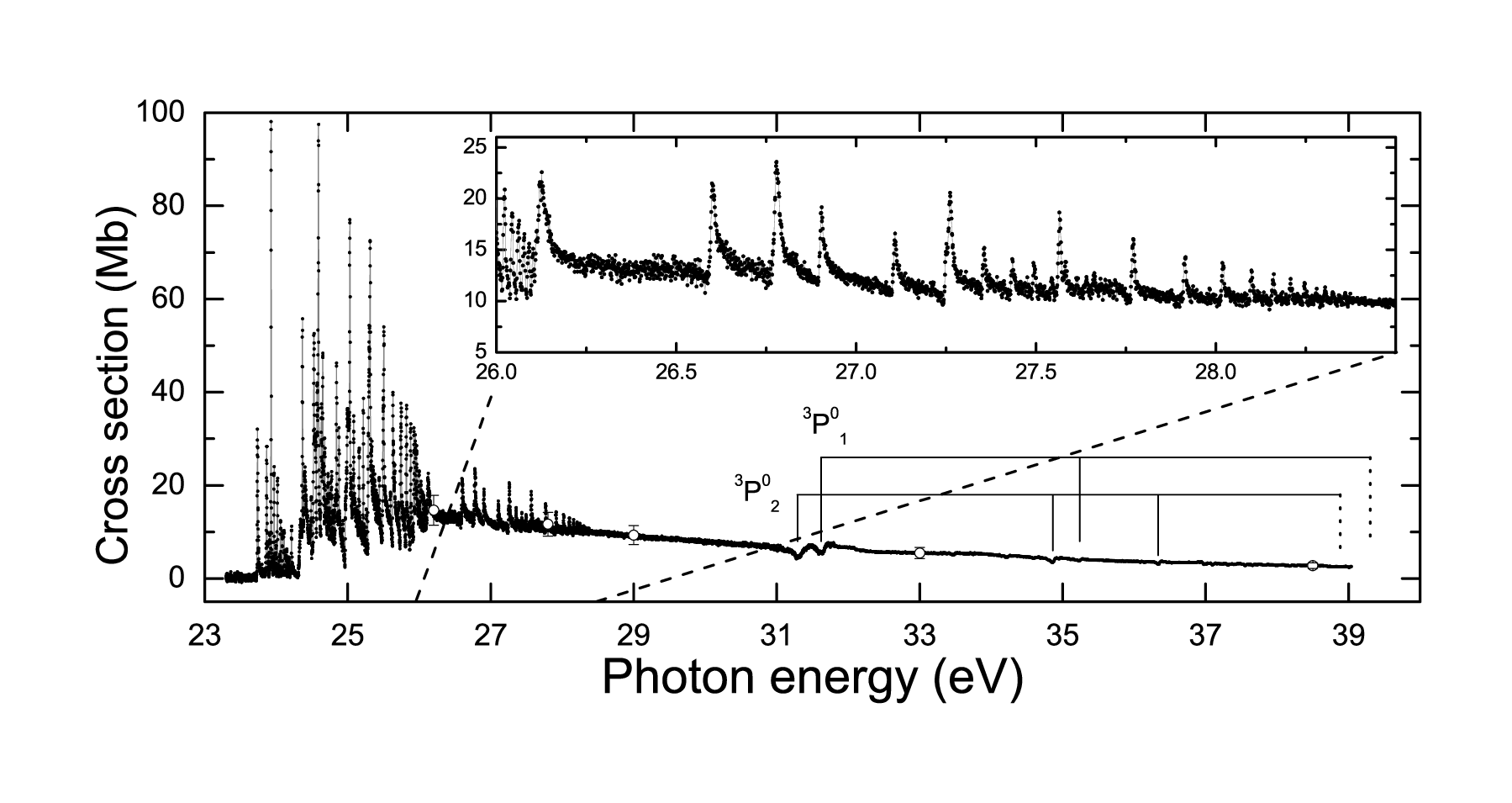}
 \caption{\label{fullrange}   An overview of single photoionization cross-sections measurements from the ALS
					 of Kr$^{+}$ ions as a function of the photon energy. 
 					The data were taken with a nominal energy resolution of 7.5 meV. 
 					Two assigned Rydberg series are indicated as vertical lines grouped by horizontal lines. 
 					The corresponding series limits $E_{\infty}$ of Equation \ref{eV} for each series are 
 					indicated by vertical-dashed lines in the end of the line groups. Labels correspond to 
					 two final states generated from initial ground state Kr$^+$($^2$P$^{\circ}_{3/2}$). Identified 
					 resonance features of this graphic are tabulated in Table \ref{WinResonances}. 
					 The inset shows the low-energy region on an expanded energy scale.
 					Assignments  at lower energies are shown in Fig. \ref{partialrange} which is a smaller range
 					of the same spectrum. The ALS absolute measurements given in Table \ref{absolutes} are shown as open circles.}
\end{center}
\end{figure*}

%
%

\section{Results}
 \subsection{Cross-section measurements}
 Absolute photoionization measurements were carried out at a number of discrete 
 photon energies where resonant structure was absent. The values of the effective 
 photoionization cross section $\sigma_{PI}$ [cm$^2$] were determined from 
 experimental parameters:

 \begin{equation}
 \sigma_{PI} = \frac{Rqe^2v_i\epsilon}{I^+I^{\gamma}\Omega \tau \Delta \int F(z)dz}
 \end{equation}

 \noindent
 where $R$ is the photoion count rate [s$^{-1}$], $q$ is the charge state of the parent
 ion, $e$ = 1.60$\times 10^{-19}$ C, \ $v_i$ is the ion beam velocity [cm$\cdot$s$^{-1}$],
 $\epsilon$ is the responsivity of the photodiode [electrons/photon], $I^{+}$ is the
 parent ion beam current [A], $I^{\gamma}$ is the photodiode current [A], $\Omega$ is
 the photoion collection efficiency, $\tau$ is the pulse transmission fraction of the
 photoion detection electronics (determined by the pulse discriminator setting), and
 $\Delta$ is the measured absolute photoion detection efficiency. The propagation
 direction of the ion beam is defined as the $z$-axis. The beam overlap integral $F(z)dz$ 
 denotes the spatial overlap of the photon and ion beams along their common interaction 
 path in units of cm$^{-1}$. At each of the three positions $z_i$ at which beam intensity 
 profiles were measured, the form factor $F(z_i)$ was determined by the following relation,

 \begin{equation}
 F(z_i) = \frac{\int \int I^+(x,y) I^{\gamma}(x,y) dxdy}{\int \int I^+(x,y)dxdy \int \int I^{\gamma}(x,y) dxdy}
 \end{equation}

\begin{figure*}
 \begin{center}
  \includegraphics[width=\textwidth]{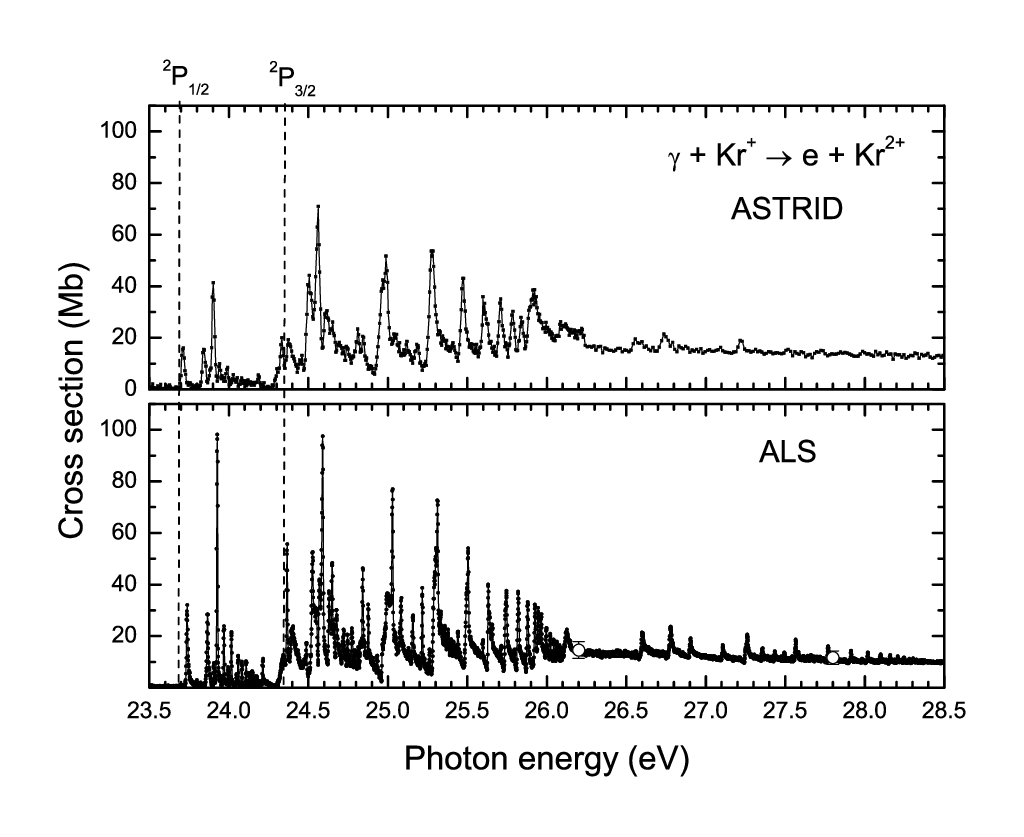}
 \caption{\label{ALSAarhuslin} Comparison of present absolute cross-section
  						measurements for single photoionization of Kr$^{+}$ taken at 7.5 meV resolution 
  						at ALS with absolute measurements of Bizau et al. ~\cite{bizau2011} measured 
						at ASTRID with a resolution of 30 meV.  The ALS absolute measurements are shown as open circles.}
\end{center}
 \end{figure*}

 The present results (see Table \ref{absolutes} for absolute values) together with measurements made at
 ASTRID \cite{bizau2011} are shown in Fig.~\ref{ALSAarhuslin}. Both
 measurements correspond to  an undetermined population-weighted sums of the cross sections for
 photoionization from the $^2$P$^{\circ}_{3/2}$ ground state and from the $^2$P$^{\circ}_{1/2}$ metastable state 
 of Kr$^+$.   Multi-configuration Dirac-Fock (MCDF) cross-sections  calculations were shown to
 compare favorably  with the ASTRID measurements, for direct (non-resonant) photoionization \cite{bizau2011}, 
assuming a statistically weighted population of levels of the $4s^24p^5$ configuration in the Kr$^+$ ion beam.  
In the present investigation we compare the higher resolution ALS measurements with state-of-the-art cross-section 
calculations performed using a Dirac-R-matrix Coulomb approximation \cite{McLaughlin2012} 
and with the earlier measurements made at ASTRID \cite{bizau2011}.
 
 The present absolute cross section measurements are systematically lower 
 than those of Bizau {\it et al.}  \cite{bizau2011}. The differences range from 
 about 15\% at 26.2 eV to a maximum difference of about 30\% 
 at 38.5 eV which is at the upper limit of their combined absolute uncertainties.  
 Even at photon energies high above the 
 ground-state threshold, the measured non-resonant cross 
 section is expected to be relatively insensitive to the admixture 
 of ground and metastable states.   
 Differences between the measured absolute cross sections 
 are attributed to a possible overestimate
 of the absolute photon flux in the present 
 experiment, caused by an undetermined small fraction of higher-order 
 radiation from the undulator that was dispersed by the monochromator 
 and recorded by the photodiode with higher sensitivity.
  
\begin{table}
\caption{\label{absolutes} Measured values of the total absolute cross section for photoionization of Kr$^{+}$
					ions from the Advanced Light Source synchrotron radiation facility in Berkeley. The total
					systematic uncertainties are estimated at a level consistent with
					90\% confidence level on statistical uncertainties~~\cite{covington2002,covington2011}.
					These absolute values are plotted as open circles 
					in Fig. \ref{fullrange},  Fig. \ref{ALSAarhuslin}, Fig. \ref{darc-astrid} and Fig. \ref{darc-astrid-als}. 
					Note that 1 Mb = 10$^{-18}$cm$^{2}$.}
\begin{ruledtabular}
\begin{tabular}{cccc}
		 Energy 	&	&  Cross section  	&	 \\
		 (eV)  	&	&  (Mb) 			 &	  \\
\hline
		  26.2  	&	&  14.6~$\pm$~4.4 	&	 \\
		  27.8  	&	&  11.7~$\pm$~3.5	&	 \\
		  29.0  	&	&  9.3~$\pm$~2.8 	&	 \\
		  33.0  	&	&  5.5~$\pm$~1.7 	&	 \\
		  38.5  	&	&  2.8~$\pm$~0.8  	&	 \\
\end{tabular}
\end{ruledtabular}
\end{table}

%
%

Figure~\ref{ALSAarhuslin} presents a comparison of  the ALS and ASTRID cross-section measurements in a photon energy
range extending about 5 eV above the ionization threshold, emphasizing the effect of photon energy resolution.

 A source of uncertainty when interpreting the measurements is the
 presence of an undetermined mixture of the $^2$P$^{\circ}_{1/2}$ metastable and the
 $^2$P$^{o}_{3/2}$ ground states in the population of the initial Kr$^{+}$ ion beam. By combining
 merged-beams and ion-trap techniques, Bizau et al. succeeded in extracting pure ground state absolute
 cross section values. Because the ion source used for the present measurements is expected to cause
some quenching of metastable states, the ground state fraction may exceed the value of 2/3 based on the relative statistical weights
of these two levels. The small differences between the two independent measurements are attributed mainly to systematic
uncertainties in both experiments. The sources of uncertainty in the present experiment are comparable in relative magnitudes to
those quoted in references~\cite{covington2002} and ~\cite{covington2011}. We also estimate an absolute energy calibration uncertainty 
of 30 meV in the ALS  spectra for this ion.

\subsection{Resonance structure}
\begin{figure*}
\begin{center}
 \includegraphics[scale=1.5,width=\textwidth]{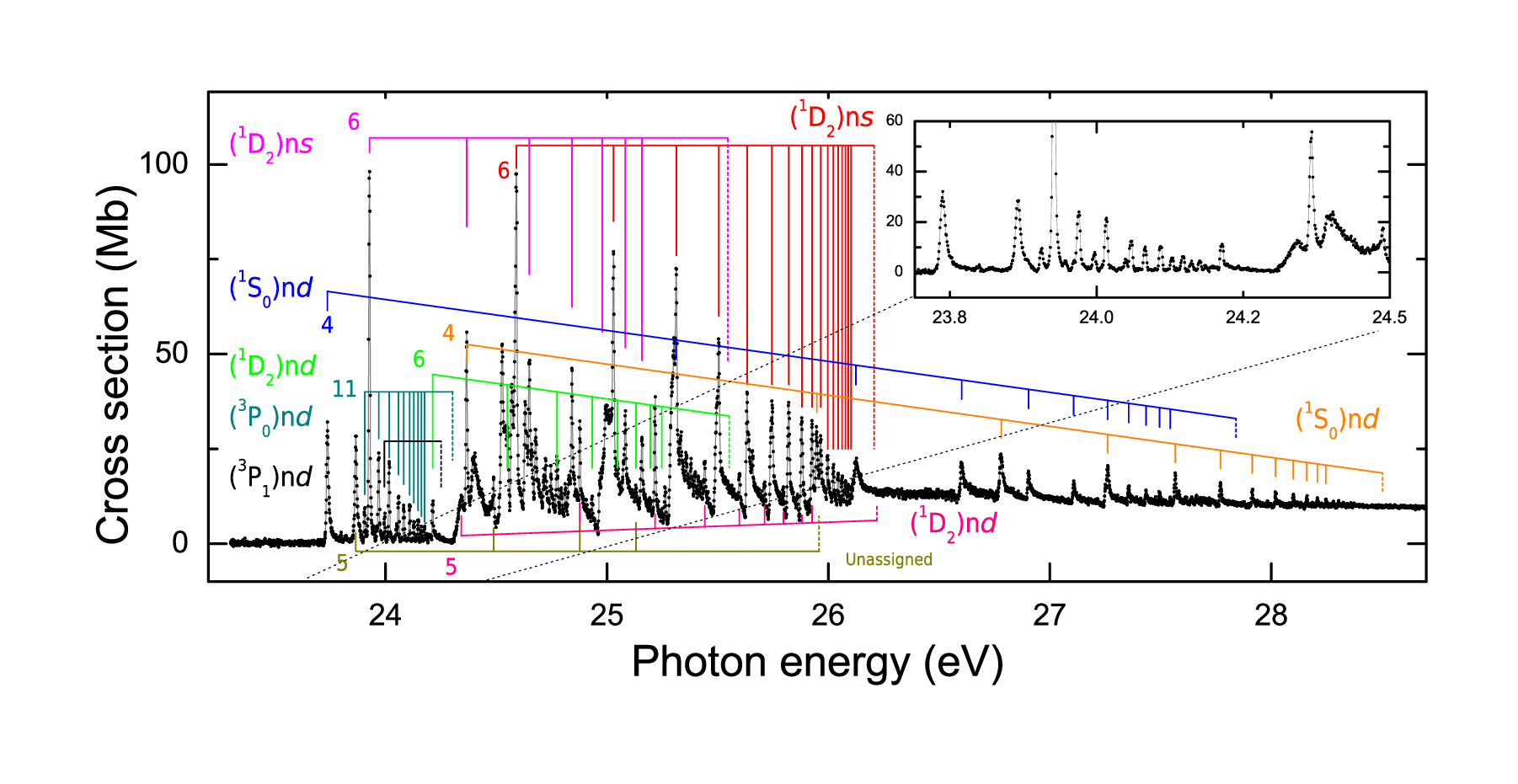}
 \caption{\label{partialrange} (Color online) Single photoionization measurements from the ALS of Kr$^{+}$ ions as a
								function of the photon energy measured with a nominal energy resolution of 7.5 meV. 
 								The assigned Rydberg series are indicated as vertical lines grouped by horizontal or inclined lines. 
								The corresponding series limits $E_{\infty}$ of Equation \ref{eV} for 
								 each series are indicated by a vertical-dashed 
								 lines in the end of the line groups. Labels to the left correspond to final states of 
 								series generated from initial metastable Kr$^+$($^2$P$^{\circ}_{1/2}$). Labels to the right 
 								correspond to final states of series generated from ground state Kr$^+$($^2$P$^{\circ}_{3/2}$). 
 								The first value of $n$ for each series is displayed close its corresponding vertical 
 								line in each group. Gaussian fit centers to the peaks are tabulated in Tables \ref{DSfromGS},
								 \ref{belowT}, and \ref{DSfromMT}. The inset shows the low-energy region on an expanded energy scale.}
\end{center}
\end{figure*}

  All the PI cross-section results determined at the ALS with a 7.5 meV spectral resolution (in the photon energy range of 23.3 eV to 39.0 eV)
 are plotted in Fig. \ref{fullrange}. In order to help guide the eye, a shorter range of the same spectrum is shown in Fig. \ref{partialrange}. 
  
 Identification of resonance structures converging to the higher energies limits 
 measured here are shown in Fig. \ref{fullrange} and in Table \ref{WinResonances}. 
 These Rydberg resonance series converge to the Kr$^{2+}$(4$s$4$p^5$) $^3$P$^{\circ}_2$, $^3$P$^{\circ}_1$ 
 series limit and appear as window resonances that are well resolved for $n =$ 3 and 4. 
 They are not well resolved for higher quantum number $n$ and the energy difference between both states is 0.412
 $\pm$ 0.030 eV, which is to be compared with the value of 0.428 eV tabulated by NIST.
 
 In Fig. \ref{partialrange} a wealth of resonance structure is shown together with the 
 proposed assignments to the most prominent features in the spectrum. In Tables \ref{WinResonances} , 
 \ref{DSfromGS}, \ref{belowT} and \ref{DSfromMT}  values of Gaussian fit centers to 
 the peaks are listed according to their assignments. The identified resonances are 
 consistent with Rydberg series originating from the ground state
 $^2$P$^{\circ}_{3/2}$ and from the metastable $^2$P$^{\circ}_{1/2}$ of the Kr$^+$ ion. These Rydberg resonance series converge respectively to
the  Kr$^{2+}$(4$s^2$4$p^4$) $^1$D$_2$, $^1$S$_0$ thresholds, and the Kr$^{2+}$(4$s$4$p^5$)  $^3$P$^{\circ}_{2,1,0}$ thresholds.
 The energy levels tabulated from references \cite{sugar1991,saloman2007} and from the NIST tabulations \cite{Ralchenko2010}
 were used as a helpful guide for the present assignments. 
 
 The resonance series identification can be made from Rydberg's formula:
 \begin{equation}\label{rydberg}
\epsilon_n  =  \epsilon_{\infty} -  \frac{{\cal~Z}^2} { \nu^{2}}    
\end{equation} 
\noindent
 where in Rydbergs $\epsilon_n$ is the transition energy, $\epsilon_{\infty}$ is the ionization potential of the excited electron  
 to the corresponding final state ($n = \infty$), i.e. the resonance series limit \cite{Seaton1983} and $n$ being the principal quantum number.
 The relationship between the principal quantum number $n$, 
the effective quantum number $\nu$ and the quantum defect $\mu$ 
for an ion of effective charge ${\cal Z}$ is given by $\nu$ = $\rm n - \mu$.
Converting all quantities to eV we can represent the Rydberg resonance series as;
\begin{equation}\label{eV}
 E_n = E_{\infty} - \frac{{\cal{Z}}^2 Ry}{(n - \mu)^2} .
\end{equation} 
\noindent
Here, $E_n$ is the resonance energy,  $E_{\infty}$ the resonance series limit, 
$\cal{Z}$ is the charge of the core (in this case $\cal{Z}$ = 2), $\mu$ is the quantum defect, being zero for a pure
 hydrogenic state, and $Ry$ is 13.6057 eV. 
 %
%
%
%

\begin{table*}                                                                                               
\caption{\label{WinResonances} 
Principal quantum numbers $n$, resonance energies (eV) and quantum 
defects $\mu$ of the higher lying Kr$^+$[$4s4p^5$ ($^3$P$^{\circ}_2$, $^3$P$^{\circ}_1$)$]nd$ series. 
The assignments are plotted in Fig. \ref{fullrange}.  
Resonance energies are calibrated to $\pm$30 meV and 
quantum defects are estimated to within an error of 20 \%.
The spectral assignments are uncertain for entries in parentheses.}

\begin{ruledtabular}      
\begin{tabular}{cccccc}                                                                                    
              			 	&          		& Rydberg series 		&         		& Rydberg series 		&             \\         
               			 	&          		& $^3$P$^{\circ}_2$     	&         		& $^3$P$^{\circ}_1$    	&              \\         
 Initial state 		 	&   $n$    		& $E_n$ (eV)     		& $\mu$   		& $E_n$ (eV)     		& $\mu$   \\  
\hline                                               																        \\
               			 	&  		  	&                				&         		&                				&               \\
 $^2$P$^{\circ}_{3/2}$ 	 &    [4$d$]     	&  31.293        			&  0.322  		&   31.622       			&  0.338  \\  
              			 	&    5     		&  34.863        			&  0.320  		&  (35.238)      			& (0.340) \\  
               				&    6     		& (36.335)       			& (0.377) 		&      -         			&     -          \\  
               				& $\cdot$  	& $\cdot$        			& $\cdot$ 		& $\cdot$        			& $\cdot$ \\
               				& $\infty$ 		& 38.882         			&         		&   39.301       			&                \\
\end{tabular}
\end{ruledtabular}                     
\end{table*}

\begin{table*}
\caption{\label{DSfromGS} Principal quantum numbers $n$, resonance energies (eV) and 
quantum defects $\mu$ of the Kr$^+[4s^24p^4 (^1D_2, ^1S_0)]ns,nd$ series. 
Resonance energies are calibrated to $\pm$30 meV and 
quantum defects are estimated to within an error of 20 \%.
The assignments are plotted in Fig. \ref{partialrange}. 
The spectral assignments are uncertain for entries in parentheses.}
\begin{ruledtabular}
\begin{tabular}{cccccc}                                                                                  
               &          & Rydberg series 	&          		& Rydberg series 	&          \\   
               &          &   $^1$D$_2$    	&          		& $^1$S$_0$      	&          \\   
 Initial state &  $n$     & $E_{n}$ (eV)   &  $\mu$   	& $E_n$ (eV)     	& $\mu$\\   
\hline                                                                            							\\
 $^2$P$^{\circ}_{3/2}$ &                	&          		&                			&          \\
               &  [6$s$]       &  24.590          	&  0.197   		&                			&          \\
               &  7       &  25.029          	&  0.200   		&                			&          \\
               &  8       &  25.313          	&  0.196   		&                			&          \\
               &  9       &  25.505          	&  0.192   		&                			&          \\
               & 10       &  25.633         	&  0.256   		&                			&          \\
               & 11       & (25.745)       	& (0.141)  		&                			&          \\
               & 12       & (25.820)       	& (0.124)  		&                			&          \\
               & 13       & (25.880)       	& (0.090)  		&                			&          \\
               & 14       & (25.926)       	& (0.066)  		&                			&          \\
               & 15       &  25.965        		& -0.005  		 &                		&          \\  
               & 16       &  25.997        		& -0.100   		&                			&          \\
               & 17       &  26.022        		& -0.011   		&                			&          \\
               & 18       &  26.043        		& -0.007   		&                			&          \\
               & 19       &  26.062        		& -0.008   		&                			&          \\
               & 20       &  26.077        		& -0.011   		&                			&          \\
               & 21       &  26.090        		& -0.012   		&                			&          \\
               & $\cdot$  & $\cdot$        	& $\cdot$ 		&                			&          \\
               & $\infty$ & 26.207         	&          		&                			&          \\ 
                                                                                  							\\
 $^2$P$^{\circ}_{3/2}$ &    		&        		&          		 	&                 \\
               &  [4$d$]        &   -            	&   -      		& (24.366)        		& (0.373)  \\
               &  5        &  24.342        		&  -0.385  		& (25.947)        		& (0.385)  \\ 
               &  6        &  24.878        		&  -0.370  		&  26.780         		&  0.379   \\ 
               &  7        &  25.217        		&  -0.370  		& (27.261)       		& (0.379)  \\ 
               &  8        &  25.441        		&  -0.370  		&  27.566         		&  0.376   \\ 
               &  9        &  25.598        		&  -0.360  		&  27.770         		&  0.378   \\ 
               & 10       &  25.712        		&  -0.358  		&  27.914         		&  0.381   \\ 
               & 11       &  25.796        		&  -0.345  		&  28.020         		&  0.383   \\ 
               & 12       & (25.880)       	& (-0.671) 	&  28.099         		&  0.381   \\ 
               & 13       & (25.926)       	& (-0.633) 	& (28.160)       		& (0.394)  \\  
               & 14       &   -            		&    -     		& (28.208)       		& (0.394)  \\
               & 15       &   -            		&    -     		& (28.247)       		& (0.418)  \\
               & $\cdot$  & $\cdot$        & $\cdot$  & $\cdot$        & $\cdot$  \\
               & $\infty$ & 26.217         &          &  28.503        &          \\
\end{tabular}
\end{ruledtabular}
\end{table*}

\begin{table*}
\caption{\label{belowT} Principal quantum numbers $n$, resonance energies (eV) and quantum 
defects $\mu$ of the Kr$^+[4s^24p^4 (^3P_2, ^3P_1)]ns$ series. 
Resonance energies are calibrated to $\pm$30 meV and 
quantum defects are estimated to within an error of 20 \%.
The assignments are plotted in Fig. \ref{partialrange}. 
The spectral assignments are uncertain for entries in parentheses.}
\begin{ruledtabular}
\begin{tabular}{cccccc}                                                                                     
               &           	& Rydberg series&          		& Rydberg series 			&        \\           
               &           	& $^{3}$P$_{0}$ &          		& $^{3}$P$_{1}$  			&        \\                                                    
 Initial state & $n$       & $E_n$ (eV)    & $\mu$    	& $E_{n}$ (eV)   			&  $\mu$ \\ 
\hline                                                                         									 \\
 $^2$P$^{\circ}_{1/2}$  &			&			&			&    			&        \\
 				& [11$s$]        	& (23.906)      	& (-0.716) 	&      -         	&   -    \\                                                    
               			& 12        		&  23.969       	&  -0.779  		&      -         	&   -    \\
               			& 13        		&  24.016       	&  -0.791  		&      -         	&   -    \\
               			& 14        		& (24.059)      	& (-0.940) 	&      -         	&   -    \\
               			& 15        		&  24.083       	&  -0.730  		&  23.996     	&  0.403 \\
               			& 16        		&  24.109       	&  -0.774  	         &      $-$       	&  $-$   \\
               			& 17        		& (24.129)      	& (-0.686) 	&      $-$       	&  $-$   \\
              			 & 18        		& (24.147)      	& (-0.707) 	&      $-$       	&  $-$   \\
               			& 19        		& (24.162)      	& (-0.659) 	&      $-$       	&  $-$   \\
               			& 20        		& (24.176)      	& (-0.731) 	&      $-$       	&  $-$   \\                
               			& $\cdot$   	& $\cdot$       	& $\cdot$  	&   $\cdot$      	& $\cdot$\\
               			& $\infty$  		& 24.303       	 &          		&    24.252      	&        \\ 
\end{tabular}
\end{ruledtabular}                                                  
\end{table*}

\begin{table*}                                                                                     
\caption{\label{DSfromMT} Principal quantum numbers $n$, resonance energies (eV) and quantum 
defects $\mu$ of the Kr$^+[4s^24p^4 (^1D_2, ^1S_0)]ns,nd$ series. 
Resonance energies are calibrated to $\pm$30 meV and 
quantum defects are estimated to within an error of 20 \%.
The assignments are plotted in Fig. \ref{partialrange}. 
The spectral assignments  are uncertain for entries in parentheses.} 
\begin{ruledtabular}  
\begin{tabular}{cccccc}                                                                                       
               &          & Rydberg series &        & Rydberg series &          \\   
               &          &$^{1}$D$_{2}$   &        & $^1$S$_0$      &          \\   
 Initial state &  $n$     & $E_{n}$ (eV)   & $\mu$  & $E_n$ (eV)     & $\mu$    \\   
\hline                                                                          \\
 $^2$P$^{\circ}_{1/2}$  & 	   		&                		&        		&                		&          \\
               			&  [6$s$]       	&  23.927        	& 0.200  $\pm$ 0.054		&                		&           \\  
               			&  7       		& (24.366)       	& (0.205)		&                		&           \\  
               			&  8       		&  24.650        	&  0.204 		&                		&           \\  
               			&  9       		&  24.842        	& 0.201  		&                		&           \\  
               			& 10       		&   -            	&   -    		&                		&           \\       
               			& 11       		&  25.082        	& 0.152  		&                		&           \\  
               			& 12       		&  25.158        	& 0.139  		&                		&           \\  
               			& $\cdot$  	& $\cdot$        	& $\cdot$		&                		&           \\
              			& $\infty$ 		& 25.545         	&       		 &                	&           \\                                                                      
               		                                                                 								    \\
 $^2$P$^{\circ}_{1/2}$  	 & 		   	&                		&        		&                	       &              \\
              			 & [4$d$]       	&   -           		&   -    		&   23.738       &  0.358   \\ 
              			 &  5       		&   -            	&   -    		&   25.312       &  0.360   \\ 
               			&  6      	 	& 24.214         	& -0.375 		&   26.124       &  0.368   \\
               			&  7      		& 24.551         	& -0.371 		&   26.602       &  0.371   \\
               			&  8       		& 24.775         	& -0.371 		&   26.904       &  0.376   \\
               			&  9       		& 24.933         	& -0.375 		&   27.108       &  0.377   \\
               			& 10       		& 25.047        	& -0.379 		&  (27.261)      & (0.303)  \\
               			& 11       		& 25.132        	& -0.363 		&   27.356       &  0.400   \\
               			& 12       		& 25.196         	& -0.349 		&  (27.435)      & (0.412)  \\
               			& 13       		& 25.248         	& -0.364 		&  (27.495)      & (0.435)  \\
               			& 14       		&   -            	&   -    		&  (27.544)      & (0.436)  \\
             			& $\cdot$ 	 	& $\cdot$        	& $\cdot$		&  $\cdot$       & $\cdot$  \\
               			& $\infty$ 		& 25.555         	&        		&   27.840       &          \\                     
\end{tabular}  
\end{ruledtabular}    
\end{table*}                                     

%

 The ionization threshold from the ground state of the Kr$^+$ ion to that of the Kr$^{2+}$ ion 
 is extremely difficult to extract from the measurements due to the presence 
 of three resonance features in that energy range:  ($^1$D$_2$)5$d$, 
 ($^1$S$_0)$4$d$ both originating from the ground state, and with ($^1$D$_2$)7$s$ originating 
 from the metastable state.

 Below the Kr$^+$($^2$P$^{\circ}_{3/2}$) ground state threshold, all transitions 
 originate from the ($^2$P$^{\circ}_{1/2}$)  metastable  of  Kr$^+$. 
 The largest structure at 23.927 eV is assigned to the transition associated with the 
 ($^1$D$_2$)6$s$ state. This series was assigned up to $n$ = 12 (see Table \ref{DSfromMT}) 
 except for the $n$ = 10 peak which is not well resolved in the spectrum. The first resonance peak
 at 23.738 eV is attributed to the transition associated with the ($^1$S$_0$)4$d$ state: resonance energies up
 to $n$ = 14 for this series are listed in Table \ref{DSfromMT}. The transition to the ($^1$S$_0$)10$d$ resonance state
 interferes with that originating from the $^2$P$^{\circ}_{3/2}$ ground state to the ($^1$S$_{0}$)7$d$ level and its assignment is uncertain,
 for this series.  We note that the intensity of  the resonance structure decreases significantly above $n$ = 12.
 
 Rydberg resonance series originating from the Kr$^{+}$(4$s^2$4$p^5$ $^2$P$^{\circ}_{1/2}$) 
 metastable state converging to the Kr$^{2+}$(4$s^2$4$p^4$ $^3$P$_0$) threshold 
 can be followed up to $n$ = 20. The intensity of its first resonance peak ($n$ = 11) is low probably due to interference with the largest 
 resonance structure below threshold. For series converging to the Kr$^{2+}$(4$s^2$4$p^4$ ~ $^3$P$_1$)  
 threshold originating from the metastable Kr$^{+}$(4$s^2$4$p^5$  $^2$P$^{\circ}_{1/2}$) state experimentally 
 we were only able to assign one resonance feature, namely the 4$s^2$4$p^4$($^3$P$_1$)15$d$ resonance state. 
 From all our resonance analysis we estimate that our quantum  defects are accurate to within 20 \%.
 
%

\begin{figure*}
\begin{center}
\includegraphics[scale=1.5,width=\textwidth,height=14.0cm]{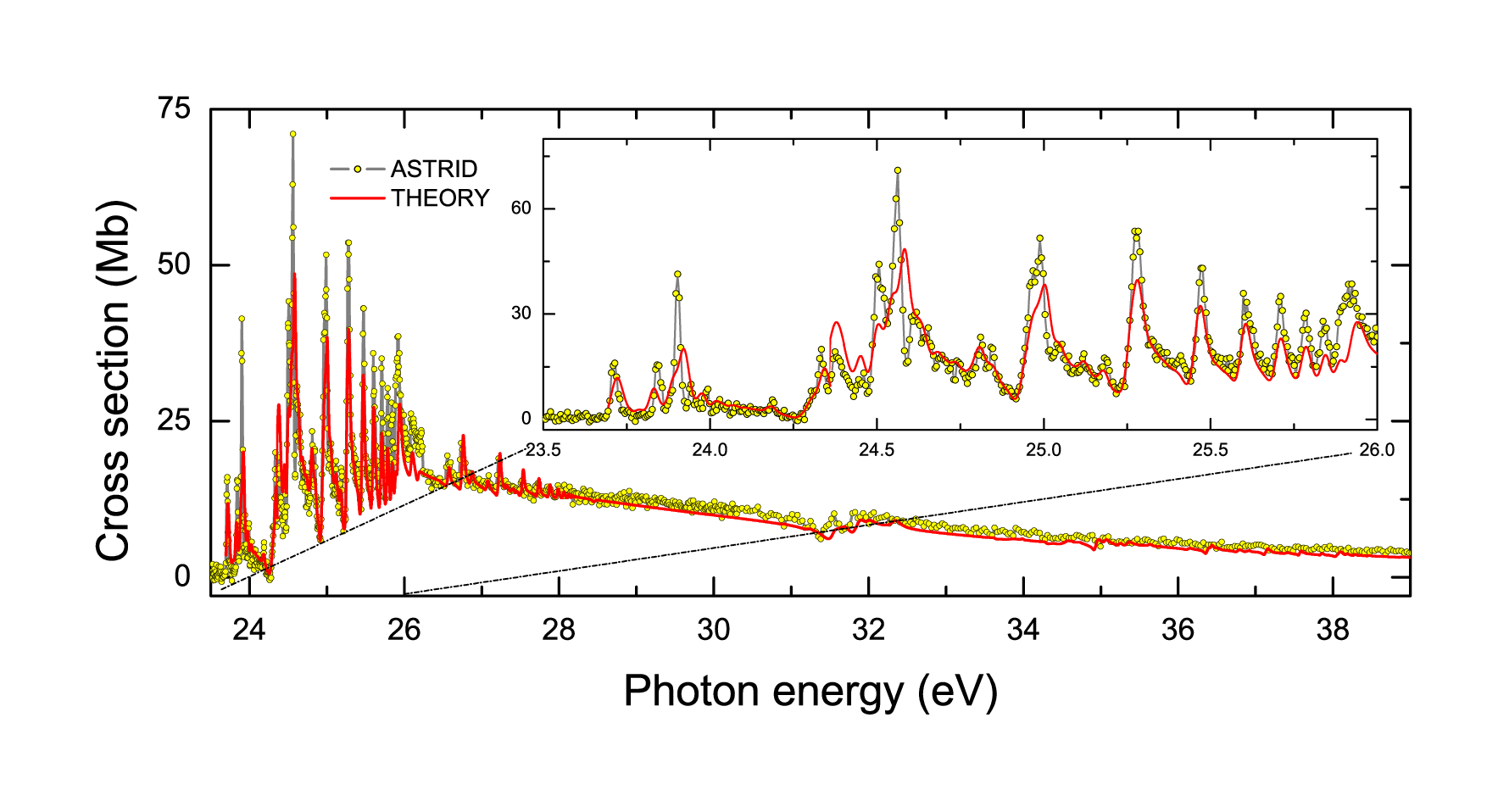}
 \caption{\label{darc-astrid} (Color online) An overview of  measurements for the absolute single photoionization measurements of Kr$^{+}$ ions as a
								function of the photon energy measured with theoretical estimates for cross sections.
								  The ASTRID/SOLEIL measurements are at a nominal energy resolution of 30 meV.
								Theoretical results were obtained from a 
								 Dirac-Coulomb R-matrix method, convoluted with the appropriate Gaussian of FWHM
								  and a statistical admixture of 2/3 ground and 1/3 metastable states.}								
\end{center}
\end{figure*}
 
 The effect of a small fraction of higher-order radiation from the undulator and dispersed by the monochromator may be more significant 
 than was originally considered at photon energies in the 20-25 eV range.
 Recently it was realized that higher-order radiation from the undulator could be considerable  \cite{Esteves2010,Esteves2011,sterling2011b,witthoeft2011}. 
 The second peak below threshold, we  tentatively speculate  the assignment to be an unidentified Rydberg series converging 
 to a limit of 25.958 eV with resolved resonance members for $n$ = 5 to 8.
 One possible assignment to this series is a Rydberg series originated from the Kr$^{+}$(4$s^2$4$p^5$~ $^2$P$^{\circ}_{3/2}$) ground state to the
Kr$^{+}$ 4$s^2$4$p^3$($^4{\rm S}^{\circ}$)n$d$($^3$D$^{\circ}$)m$g$ transitions, caused by second order radiation in the photon beam. However,
 this assignment is unclear because its energy limit is shifted by 0.066 eV from the NIST tabulated value.
 
 Above the ground-state ionisation threshold, Rydberg resonance series originate from both the initial metastable Kr$^+$($^2$P$^{\circ}_{1/2}$) 
 and the ground state Kr$^+$($^2$P$^{\circ}_{3/2}$) and converge respectively  to the Kr$^{2+}$(4$s^2$4$p^4$ $^1$D$_2$,  $^1$S$_0$) thresholds. 
 For example, in Fig. \ref{partialrange}, two series converging to the Kr$^{2+}$(4$s^2$4$p^4$ $^{1}$S$_0$) threshold are indicated by 
 inclined lines. The ionization potential difference of these two series is 0.663 eV which is 
 consistent with  the value of 0.666 eV tabulated by NIST \cite{Ralchenko2010}. 

 The quantum defects $\mu$ of each of the Rydberg resonance series 4$s^2$4$p^4$($^1$S$_0$)n$d$  
 converging to Kr$^{2+}$(4$s^2$4$p^4$ $^1$S$_0$) threshold originating  from 
 the $^2$P$^{\circ}_{3/2}$ initial ground state respectively show a nearly constant value.  
 Any changes we attribute to the interference with the neighboring resonance structures. They
 have an average value of 0.385 $\pm$ 0.08. Similarly for the quantum defects of the 
 corresponding Rydberg resonance series 4$s^2$4$p^4$($^1$S$_0$)n$d$  originating 
 from the $^2$P$^{\circ}_{1/2}$ metastable state show a monotonically increasing value with an average of 0.381 $\pm$ 0.08.  These compare 
 favorably with the average value of 0.331 from the work of McLaughlin and Ballance \cite{McLaughlin2012}. 
 The Rydberg resonance series 4$s^2$4$p^4$($^1$D$_2$)$ns$ converging  to the ($^1$D$_2$) threshold 
 have quantum defects $\mu$ $\sim$ of  0.2 $\pm$ 0.04 or less 
 consistent with current average estimates of 0.19.
 Quantum defect  values  ($\mu$) for Rydberg resonance series 4$s^2$4$p^4$($^1$D$_2$)$ns$,$md$  
 as  measured by Ru$\check{\mbox{s}}\check{\mbox{c}}$\'ic et al. \cite{berko1984}
 for Kr$^+$-like atomic Bromine converging to ($^1$D$_2$) threshold are consistent with the
 present values. For the case of the quantum defects values ($\mu$ ) reported by  Bizau et al. \cite{bizau2011},
 the agreement is qualitatively consistent for states converging to the $^1$S$_0$ threshold. 
 
 Some of the features found  in the spectrum could not be assigned, in particular, the large peak at 24.526 eV  and the structures
 in the range 24.678 eV to 24.746 eV remain as yet unidentified.

%

\begin{figure*}
\begin{center}
\includegraphics[scale=1.5,width=\textwidth,height=14.0cm]{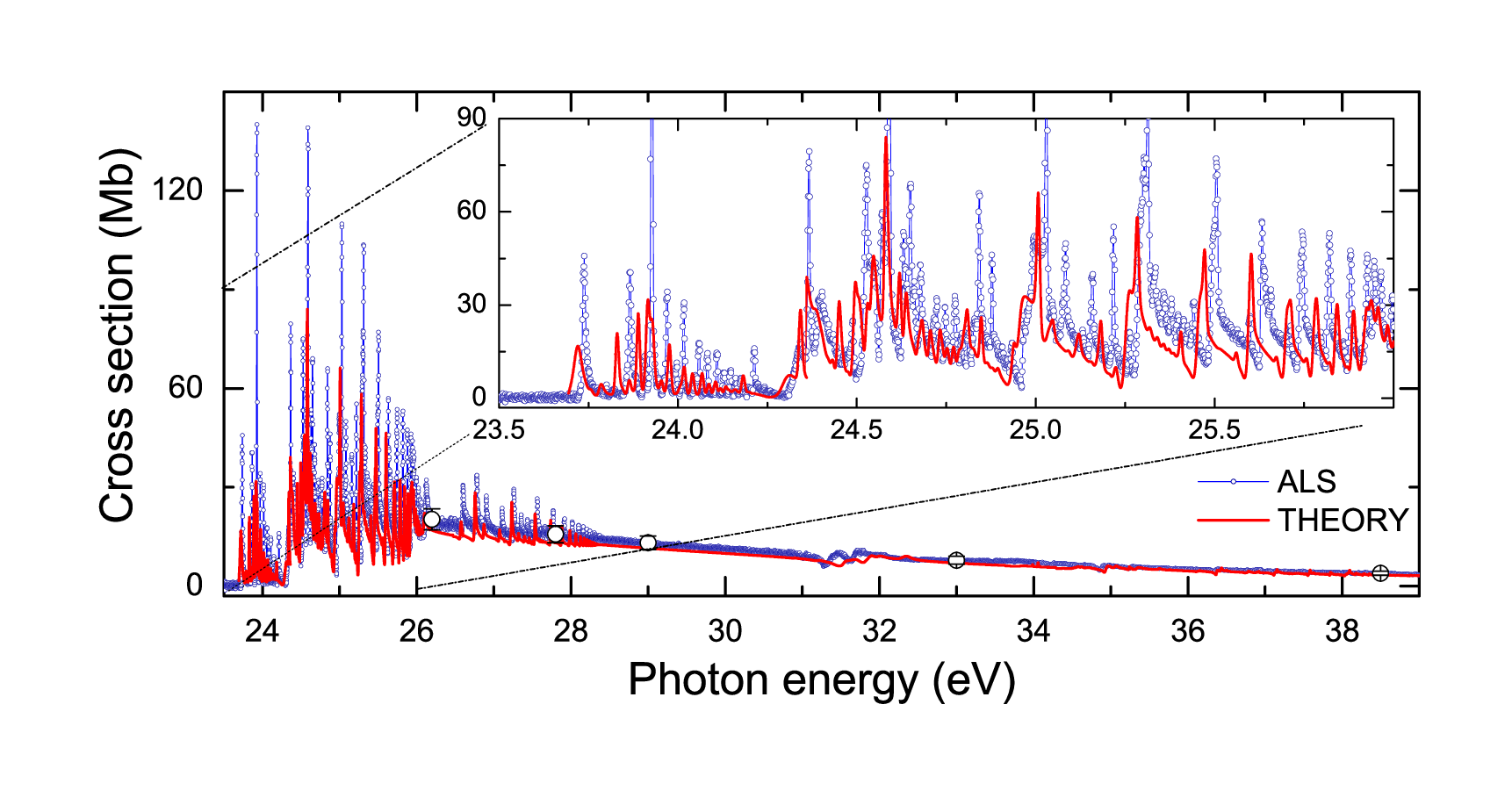}
 \caption{\label{darc-astrid-als} (Color online) An overview of measurements for the absolute single photoionization measurements of Kr$^{+}$ ions as a
								function of the photon energy measured with theoretical estimates for cross sections.
								  The ALS measurements are at a nominal energy resolution of 7.5 meV and normalised to the ASTRID/SOLEIL
								  measurements at 26.5025 eV.
								Theoretical results were obtained from a 
								 Dirac-Coulomb R-matrix method, convoluted with the appropriate Gaussian of FWHM
								  and a statistical admixture of 2/3 ground and 1/3 metastable states.}								
\end{center}
\end{figure*}

\section{Discussion}
The higher-resolution ALS experimental PI cross-section measurements were taken at 7.5 meV FWHM  
compared  to the ASTRID/SOLEIL measurements obtained at the lower spectral energy resolution of 30 meV FWHM. 
To simulate the ALS experimental conditions the DARC theoretical PI cross sections 
were convoluted with a Gaussian of 7.5 meV FWHM and statistically averaged.
In prior comparisons of the DARC PI cross sections with the ASTRID/SOLEIL 
 measurements taken at a spectral resolution of 30 meV FWHM \cite{McLaughlin2012}, 
the hypothesis of statistical population was justified since the
life time is huge (0.35 seconds for the case of Kr$^\mathrm{+}$) if compared to the beam transport
times (microseconds). Excited $\rm ^2P^{\circ}_{1/2}$ metastable Kr$^\mathrm{+}$ ions produced
either in an ion source or ion trap rarely collide with surfaces or residual
gas before photoionization takes place. Statistical population of the ions seems consistent with their measurements
both accelerating ions from an ion source and using an ion trap without delay \cite{bizau2011}.  For the case of Kr$^\mathrm{+}$ ions 
studied here we note that the theoretical results obtained from the DARC codes for  the resonance  positions and quantum 
defects are in reasonable agreement with the experimental measurements made at ASTRID and SOLEIL ~\cite{bizau2011}
where a statistical weighting of the initial states was assumed \cite{McLaughlin2012}.
Taking this into account, Fig. \ref{darc-astrid} indicates, for the spectral range investigated
 the magnitude and shape of the absolute cross sections are in excellent agreement 
with the previous ASTRID/SOLEIL experimental data taken at the lower resolution of 30 meV,
 along with the resonance energy positions.  The same photon energy range  is shown in Fig. \ref{darc-astrid-als} 
 where the prominent Rydberg resonance series are highlighted and compared 
 with the higher resolution ALS measurements. 
 Here again we have excellent agreement between theory and experiment.
 
 The integrated oscillator strengths $f$ of the spectra, calculated using \cite{Fano1968}
 \begin{equation}
 f =9.11 \times 10^{-3} \int \sigma (h\nu) {\rm d} h \nu
 \end{equation}
 gives a value of 1.40 $\pm$ 0.10 from the ASTRID/SOLEIL measurements of Bizau and co-workers \cite{bizau2011}. 
 From the DARC R-matrix PI cross section calculations we obtain a value of 1.34,  
in excellent agreement with the previous lower resolution ASTRID/SOLEIL measurements. 
The present absolute measurements from the ALS yield an oscillator strength of 1.10 $\pm$ 0.25 over the same energy range. 
As previously noted, this difference is attributed to an over-estimate of the photon flux due to a small fraction of higher-order radiation in the photon beam.

\section{Summary}
Photoionization cross sections were measured for a statistical admixture of ground
 state and metastable Kr$^+$ ions in the photon energy range 23.3 eV to 39.0 eV with a
spectral energy resolution of 7.5 meV.   The quantum defects for each of the Rydberg series 
 are estimated to have an uncertainty of 20 \% and resonance energies an uncertainty of $\pm$30 meV. 
 The dominant features were spectroscopically assigned to Rydberg series and quantum
 defects were determined. The absolute measurements reported here are in agreement with recent
 measurements by Bizau et al.~\cite{bizau2011} within their combined systematic uncertainties.
 The higher energy resolution of the present measurements permitted additional resonance structures
 to be identified and additional spectroscopic assignments to be made. Comparisons of the statistically averaged
PI cross section calculations from the Dirac-Atomic-R-matrix (DARC) codes with measurements, particularly in the resonant region 
are in excellent agreement with the ALS higher resolution measurements and with the lower resolution measurements from ASTRID.
Overall excellent agreement is seen between experiment and theory  for 
 this complicated trans-Fe element for the entire photon energy investigated.
 The present data render much higher resolution data than previous measurements \cite{bizau2011},
 providing better assignments to spectroscopic features and have been benchmarked 
 against state-of-the-art PI calculations using the Dirac-Atomic-R-matrix-codes (DARC).
 The photoionization cross-sections from the present study are suitable to be included into
state-of-the-art photoionization modelling codes such as CLOUDY  \cite{ferland1998,ferland2003}, 
XSTAR \cite{kallman2001} and ATOMDB \cite{Foster2012}  that are used to numerically
simulate the thermal and ionization structure of ionized astrophysical nebulae.

%
%
%
%

\begin{acknowledgments}
We thank Professor J. M. Bizau from the Universit\'e Paris-Sud, France
for the published merged-beam data from his group to compare with the present work.
The experimental work was supported by the Office of Basic Energy
Sciences, Chemical Sciences, Geosciences and Energy Biosciences
Division, of the U.S. Department of Energy under grants
DE-FG03-00ER14787 and DE-FG02-03ER15424 with the University of
Nevada, Reno; by the Nevada DOE/EPSCoR Program in Chemical Physics
and by CONACyT, Cuernavaca, M\'{e}xico. 
Grants DGAPA UNAM-IN 113010, through ICF-UNAM, Cuernavaca, M\'exico are acknowledged by
I. \'{A}., C. C. and G. H.,  M. M. S'A.  acknowledges support from CNPq (Brazil).
C. P. B.  was supported by US Department of Energy (DoE)
grants  through Auburn University.
B. M. McL. acknowledges support by the US
National Science Foundation under the visitors program through a grant to ITAMP
at the Harvard-Smithsonian Center for Astrophysics.  
The computational work was performed at the National Energy Research Scientific
Computing Center in Oakland, CA, USA and on the 
Kraken XT5 facility at the National Institute for Computational Science (NICS) in Knoxville, TN, USA.
 The Kraken XT5 facility is a resource of the Extreme Science and Engineering Discovery Environment (XSEDE), 
which is supported by National Science Foundation grant number OCI-1053575.
The Advanced Light Source  is supported by the Director, 
Office of Science, Office of Basic Energy Sciences, 
of the US Department of Energy under Contract No. DE-AC02-05CH11231.

\end{acknowledgments}
%
%
%
%
\bibliographystyle{apsrev4-1}
\bibliography{krplus}

\end{document}